\documentclass[pre,twocolumn,groupedaddress,showpacs,showkeys,amsmath]{revtex4}

\usepackage{graphicx}
\usepackage{dcolumn}
\usepackage{bm}

\begin{document}


\title{Typical and large-deviation properties of minimum-energy paths 
on disordered hierarchical lattices}

\author{O. Melchert}
\email{oliver.melchert@uni-oldenburg.de}
\author{A. K. Hartmann}
\email{alexander.hartmann@uni-oldenburg.de}
\affiliation{
Institut f\"ur Physik, Universit\"at Oldenburg, Carl-von-Ossietzky Strasse, 26111 Oldenburg, Germany\\
}

\date{\today}


\begin{abstract}
We perform numerical simulations to study the optimal path problem 
on disordered hierarchical graphs with effective dimension 
$d_{\rm eff}\approx2.32$. 
Therein, edge energies
are drawn from a disorder distribution that allows for positive and negative 
energies. This induces a behavior which is fundamentally
different from the case where all energies are positive, only.
Upon changing the subtleties of the distribution, the scaling of the
minimum energy path length 
exhibits a transition from self-affine to self-similar.
We analyze the precise scaling of the path length and the associated
ground-state energy fluctuations in the vincinity of the disorder critical point,
using a decimation procedure for huge graphs. 
Further, using an importance sampling procedure in the disorder we compute
the negative-energy tails of the ground-state energy distribution up to 
$12$ standard deviations away from its mean.
We find that the asymptotic behavior of the negative-energy tail is 
in agreement with a Tracy-Widom distribution. Further, the characteristic
scaling of the tail can be related to the ground-state energy flucutations,
similar as for the directed polymer in a random medium.
\end{abstract} 

\pacs{02.60.Pn,05.10.Ln,64.60.De}
\keywords{Disordered hierarchical lattice, minimum-weight path, large-deviation properties}
\maketitle

\section{Introduction \label{sect:introduction}}

Many problems in physics and computer science can 
conveniently be modeled using graphs. Thereby it is often
inevitable to
assign attributes to the edges that assist in specifying the problem under 
consideration. E.g., weighted graphs, where a weight,
 is associated 
with each edge, might be used to model disordered environments. 
For a given weighted graph the \emph{minimum-energy path} (MWP) problem
refers to the paradigmatic optimization problem of finding a simple (i.e.\ loopless) path, 
connecting two distinguished nodes of the graph, along which the sum of the 
edge weights is minimal.
The MWP problem quite naturally lends itself to study a multitude of 
lattice-path models in the context of disordered systems.
In this regard, it has proven
to be useful in order to characterize, e.g.,
linear polymers in random media \cite{kremer1981,kardar1987,derrida1990,grassberger1993,parshani2009}, 
domain wall excitations in disordered environments such as spin glasses \cite{cieplak1994,melchert2007,melchert2011} and 
the solid-on-solid model \cite{schwarz2009}. Due to this relation
to physical problems, the weight will be denoted as \emph{energy}
in the following. 
If the disorder is drawn from a distribution that allows for nonegative 
edge energy only, as for the canonical ``directed polymer in a random 
medium'' (DPRM), the groundstate configuration of the polymer can be 
computed efficiently using Dijkstra's algorithm \cite{schwartz1998,clrs2001}.
However, if the disorder distribution allows for edge-energies 
of either sign, as for the problem of finding a minimum energy 
domain wall in $2D$ Ising spin glasses \cite{melchert2007,melchert2009,melchert2011}
(given that there is no closed path with a negative energy) or more
generally for the 
\emph{negative-weight percolation} (NWP) problem 
\cite{melchert2008,apolo2009,melchert2010a,melchert2011b,Norrenbrock2012,claussen2012}, the 
solution of the MWP problem requires a nontrivial transformation to an 
auxiliary 
minimum-weight perfect matching problem \cite{ahuja1993}.
Furthermore, the properties of MWPs with negative edges are
fundamentally different from the case where all edge energies
are non-negative \cite{melchert2007,melchert2009,melchert2011}.

\begin{figure}[t!]
\centerline{
\includegraphics[width=1.0\linewidth]{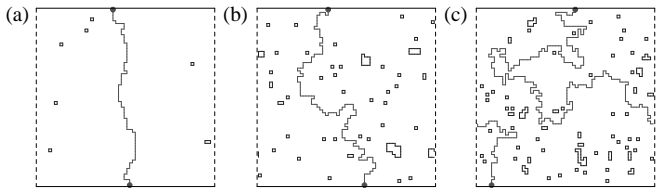}}
\caption{
Samples of minimum-energy configurations consisiting of 
one path (forced to span the lattice along the direction 
with the free boundaries) and a set of loops for a $2D$ square lattice 
with side length $L\!=\!64$ and periodic boundaries
in the horizontal direction.
The snapshots relate to different values of the
disorder parameter $\rho$, where (a) $\rho<\rho_c$, 
(b) $\rho\approx \rho_c$, and, (c) $\rho>\rho_c$.
In the limit of large system sizes and above the critical point 
$\rho_c$, paths might span the lattice along the direction 
with the periodic boundaries.
\label{fig:samples2D}}
\end{figure}

To specify the NWP more precisely, one considers, say,  
a regular $d=2$ square lattice 
graph with side length $L$ and free boundaries in one direction, periodic boundaries in the remaining
direction, and energies drawn from 
a distribution that allows for edge energies of either sign. The details
of the energy distribution are controlled by a tunable disorder 
parameter.
For a given realization of the disorder one might be interested in,
say, an agent ``harvesting'' the negative energies 
(seeing as a negative cost, i.e.,
a resource, e.g., an energy) along a freely adjustable path between two given points.
This means the walker might have to use  edges with positives energies as
well, i.e., spend some amount of the resource. In addition
some resources might be harvested, in parallel or in competition
to the walker, by other walkers which are not restricted
to walk between two given endpoints. These other walkers are only present
for walks where the amount of the harvested  resource is larger 
than the amount of
the resource spent. Mathematically this means we
consider
configurations consisting of a single path and a set of loops, 
i.e.\ closed paths, 
such that the total sum of the energies assigned to the edges
that build up the path and the loops attains a  minimum. 
As an additional optimization constraint the path might be forced to span the
lattice along the direction with the free boundaries. 
This means, the walker covers a large fraction of the lattice
which allows him to maximize the amount of the harvested resource.
Further the path and the loops are simple and 
are not allowed to intersect each other.
Therefore, they exhibit an ``excluded volume''
quite similar to usual self avoiding walks (SAWs) \cite{stauffer1994}.
A pivotal observation is that the NWP model features a disorder driven, geometric phase transition,
signaled by the emergence of paths that span the lattice along the direction 
with the periodic boundary conditions.
In this regard, depending on the disorder parameter, one can identify two distinct
scaling regimes: 
(i) a phase where the paths tend to be short in length, displaying a self-affine 
scaling with system size, see Fig.\ \ref{fig:samples2D}(a), and,
(ii) a phase where the paths tend to be long and exhibit a self-similar scaling, see Figs.\ \ref{fig:samples2D}(b),(c).
From the previous analyses for $d=2$ we found that right at the critical point, the paths are self-similar 
with a fractal dimension $d_f=1.268(1)$, see Refs.\ \cite{melchert2008,Norrenbrock2012}.

Here, we study a particular MWP problem (which closely resembles the NWP problem) in a
 Migdal-Kadanoff-like renormalization group scheme on hierarchical lattice graphs 
(constructed using a ``Wheatstone bridge'' elementary cell, see Fig.\ \ref{fig1})
with an effective dimension $d_{\rm eff}\approx2.32$ \cite{cieplak1995,Salmon2010,Teodoro2010}, 
where an exact decimation 
procedure can be used to analyze huge graphs \cite{bray1987,boettcher2003}. 
We address the critical behavior of the MWP in this setup by monitoring
observables related to the path energy and path length. The subtleties of the 
construction procedure that leads to hierarchical lattices with effective
dimension $d_{\rm eff}\approx2.32$ even allows to probe the transition 
from the self-affine to self-similar scaling of the path length, as 
observed for the NWP problem on hypercubic lattice graphs \cite{melchert2008,melchert2010a}.

Similar to previous studies of minimum-energy path problems on hierarchical 
graphs \cite{Derrida1989,derrida1990,Doussal1991,Devillard1993,Cao1993,cieplak1994,cieplak1995,Shussman1995,monthus2008} and regular lattices \cite{kremer1981,kardar1987,schwartz1998,Hansen2004,buldyrev2006,parshani2009}, we here consider the finite-size scaling 
of the length and energy of the paths as well as the 
associated energy fluctuations.
We further complement the simple sampling (SiSa) estimates of the 
path-energy distribution by an importance
sampling (ImSa) procedure in the disorder 
\cite{align2002,rare-graphs2004,koerner2006,monthus2006,align_long2007}, 
allowing to resolve the respective distributions up to 12 
standard deviations away from its mean.
In this regard it is found that the asymptotic behavior of the negative-energy tail is 
in agreement with a Tracy-Widom distribution. Further, the characteristic
scaling of the tail can be related to the path energy fluctuations,
similar as for the directed polymer in a random medium \cite{Doussal1991}.
Note that, apart from the disorder distribution, the MWP problem considered here is similar to the 
optimal path problem on hierarchical lattices as studied in Ref.\ \cite{cieplak1995} (therein, the authors considered
a uniform distribution of nonegative edge-energies, only).

Above we pointed out that the MWP problem studied here closely resembles the 
NWP problem. At this point we would like to point out the major similarities
and differences of both models: similar to the NWP problem, in the 
MWP problem the sum of energies of edges that build up a path is object to
minimization. As a major difference note that the path found in the context of
the NWP problem is not necessarily the (absolute) minimum-energy path.
The reason is that in the NWP problem, a global minimum of the energy for a 
single path plus a (possibly empty) set of loops (all with negative energy) 
is searched for, see Fig.\ \ref{fig:samples2D}.
This is in contrast to the MWP problem on hierarchical lattices, where a 
particular decimation scheme, see sect.\ \ref{sect:model}, allows to obtain 
a truly minimum-energy path in a framework where no loops are considered. 
However, also note that at the critical point of the NWP 
model, see Fig.\ \ref{fig:samples2D}(b), the ``additional'' loops are small and
resemble a rather dilute ``gas'' of loops which are unlikely to affect the statistics
of the path that spans the system in between the free boundaries.
Hence, in the vicinity of the critical point of the model we expect the 
MWP problem studied here to provide a reasonable approximation to the 
path in the NWP problem.

The remainder of the presented article is organized as follows.
In section \ref{sect:model}, we explain the construction procedure 
to obtain the hierarchical lattice graphs and we outline the pool 
method used compute the properties of the paths for huge graphs. 
In section \ref{sect:results}, we list the results of 
our numerical simulations in terms of which we locate the 
self-affine to self-similar transition of the path length and 
where we put under scrutiny the path-energy distribution. 
In section \ref{sect:conclusions} we 
conclude with a summary.


\section{Model and Algorithm\label{sect:model}}
\begin{figure}[t!]
\centerline{
\includegraphics[width=1.0\linewidth]{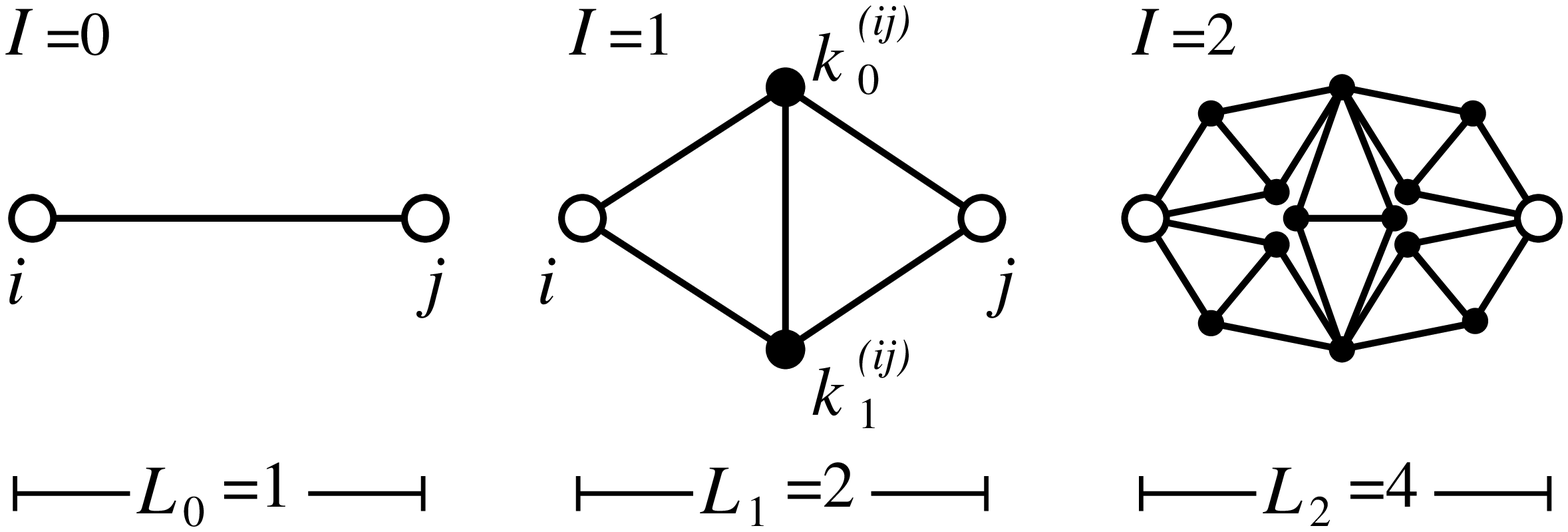}}
\caption{Illustration of the first two iteration steps in the construction 
procedure to obtain the hierarchical graphs for which the presented 
study is carried out.
The linear extend $L_I$ of the graphs $G_I$ is also indicated.
As explained in the text, the resulting graphs have an effective 
dimension $d_{\rm eff}\approx 2.32$.
\label{fig1}}
\end{figure}  

The hierarchical lattices considered in the remainder of
the presented article can be constructed using a simple 
deterministic rule. 
This rule specifies how the individual edges in a graph 
at a given iteration step $I$ need to be transformed 
in order to obtain a graph at iteration step $I+1$.
Let $G_I=(\mathcal{V}_I,\mathcal{E}_I)$ denote a hierarchical graph that consists
of a set of nodes $i\in \mathcal{V}_I$ and a set 
$\mathcal{E}_I \subset \mathcal{V}_I^{(2)}$ of undirected edges
$e=\{i,j\}\in \mathcal{E}_I$. The number of 
edges is given by $M_I=|\mathcal{E}_I|$.
The transformation in order to proceed from $G_I$ to $G_{I+1}$ reads 
as follows:
each edge $e=\{i,j\}\in \mathcal{E}_I$ is replaced by a subgraph $G^\prime$ consisting 
of four nodes $\{i,k_0^{(ij)},k_1^{(ij)},j\}$ (therefore the set of 
nodes needs to be amended by two nodes $k_0^{(ij)}$ and $k_1^{(ij)}$)
and five edges $\{\{i,k_0^{(ij)}\},\{i,k_1^{(ij)}\},\{k_0^{(ij)},k_1^{(ij)}\},\{k_0^{(ij)},j\},\{k_1^{(ij)},j\}\}$.
The nodes $i$ and $j$ are referred to as the \emph{terminal nodes} of the subgraph.
After the transformation is completed the number of edges increased to $M_{I+1}=5 \times M_I$,
and the linear extension of the graph has doubled, i.e.\ $L_{I+1}=2 \times L_{I}$.
At $I=0$ the construction procedure is started with a single 
edge, meaning that $M_0=1$ and $L_0=1$. Hence, $M_I=5^{I}$ and
$L_I=2^I$. From the increase of the 
number of edges $M_I$ as a function of the linear extension $L_I$
of the graphs according to $M_I=2^{I \log_2(5)}=L_I^{d_{\rm eff}}$ it is possible 
to obtain the effective (fractal) dimension of the hierarchical 
lattices as $d_{\rm eff}=\log_2(5)\approx 2.32$.
The construction procedure is illustrated in Fig.\ \ref{fig1}, where, 
starting with a single edge at $I=0$, the two steps $G_0 \to G_1 \to G_2$ 
are shown explicitly.
Finally, a \emph{path} is represented by an ordered set of edges. E.g., regarding
the subgraph $G^\prime$, a possible path that connects its terminal nodes
$i$ and $j$ reads $p=(\{i,k^{(ij)}_0\},\{k_0^{(ij)},k_1^{(ij)}\},\{k_0^{(ij)},j\})$.

\begin{figure}[t!]
\centerline{
\includegraphics[width=1.\linewidth]{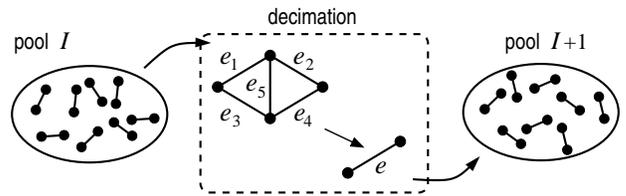}}
\caption{
Illustration of the decimation procedure used to fill
the pools. Repeatedly, five edges $e_1 \ldots e_5$ are 
randomly picked from pool $\mathcal{P}_I$ in order to 
construct a five-edge subgraph $G^\prime$ which
is in turn decimated to a single edge $e$ (see text) and added to 
pool $\mathcal{P}_{I+1}$. 
\label{fig:poolMethod}}
\end{figure}

The minimum-energy path problem we address here reads as 
follows. Let $s$, $t$ denote the endnodes of the single 
edge at $I=0$. Perform a number of $I_{\rm max}$ iteration
steps to yield a hierarchical graph $G_{I_{\rm max}}$ and
assign a random energy to each edge, drawn from a given disorder
distribution. Finally, compute a minimum 
energy $s$-$t$ path for the graph $G_{I_{\rm max}}$.
The precise topology of the resulting path depends on the particular realization of 
the disorder and has length $\ell \in [2^{I_{\rm max}},3^{I_{\rm max}}]$.
Bear in mind that in order to compute one such path, a graph with $M_{I_{\rm max}}=5^{I_{\rm max}}$
edges needs to be constructed. Consequently, a number of $M_{I_{\rm max}}$
random deviates need to be drawn from the disorder distribution.
A more efficient way to sample minimum energy $s$-$t$ paths 
for the case of hierarchical graphs at large values of $I$ 
is provided by the \emph{pool method} \cite{bray1987,boettcher2003}.
Therein one maintains a set of $I_{\rm max}$ pools $\mathcal{P}_I$
of (effective) edges, 
with $I=0\ldots I_{\rm max}-1$. The number
of edges in each pool is the same and
is denoted by $N$. 
An individual edge carries two attributes $e=(E,\ell)$, 
where $E$ denotes the energy and $\ell$ the length of a path associated with the edge.
At $I=0$ the edges are initialized with $\ell=1$ and the energies $E$
are drawn from a specified disorder distribution $P_0(E)$, signifying the ``single edge level''.
In order to 
proceed from pool $\mathcal{P}_I$ to $\mathcal{P}_{I+1}$, the following 
three-step \emph{decimation procedure}, sketched in Fig.\ \ref{fig:poolMethod}, 
has to be repeated $N$ times:
\begin{enumerate}
\item[(i)] pick five edges $e_1 \ldots e_5$ at random from pool $\mathcal{P}_I$.
Combine these to form a subgraph $G^\prime$ as explained earlier.
\item[(ii)] from the four distinct paths that connect the terminal nodes of $G^\prime$, 
determine the minimal energy path $p^\star$, i.e.\ the path 
$p^\star \in \{(e_1,e_2),(e_3,e_4),(e_1,e_5,e_4),(e_3,e_5,e_2)\}$ 
for which $E^\star \equiv \sum_{e\in p^\star} E(e) \stackrel{!}{=} {\rm min}$. 
Correspondingly, the length of the path reads $\ell^\star =\sum_{e\in p^\star} \ell(e)$.
\item[(iii)] set up a new edge having attributes $e=(E^\star,\ell^\star)$ and add it to pool $\mathcal{P}_{I+1}$. 
\end{enumerate}

After $\mathcal{P}_0$ has been initialized, all pools up to $I=I_{\rm max}$ might be 
filled in this manner. Note that an edge $e \in \mathcal{P}_I$ effectively 
corresponds to a hierarchical graph $G_I$, i.e.\ it has a ``hidden'' substructure 
that allows to represent a minimum-energy path with length $\ell \in [2^I,3^I]$.
The attributes of the edge encode the characteristics of the respective path, i.e.\
its energy $E$ and length $\ell$. A pool $\mathcal{P}_I$ thus consists of $N$ instances
of minimum-energy paths for hierarchical graphs at iteration step $I$.
Thereby, the computational resources needed to fill a pool stay constant as $I$ increases. 

\begin{figure}[t!]
\centerline{\includegraphics[width=1.\linewidth]{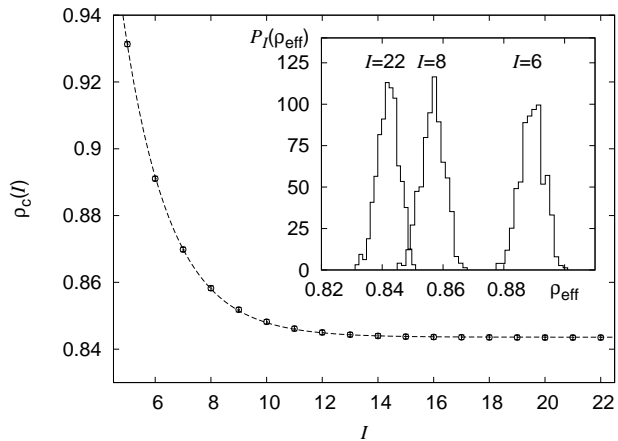}}
\caption{
Results for the estimation of the critical point $\rho_c$ using
the secant method.
The main plot shows the scaling of the iteration step dependent critical points
$\rho_{c}(I)$. The 
dashed line is a fit to the scaling form $\rho_c =\rho_c + a 2^{-I b}$
in the interval $I\in [5,22]$, resulting in the estimates
$\rho_c=0.8436(2)$, $a=O(1)$, and $b=0.866(4)$. 
The inset illustrates the probability density function $P(\rho_{\rm eff})$ of 
the effective critical values $\rho_{\rm eff}$ for $I=6,8,22$ where 
a number of $256$ independent estimates were considered. 
\label{fig:secantMethod}}
\end{figure}  

Further, each pool specifies its own distributions $P_I(E)$ and $P_I(\ell)$ of path energies and 
path lengths, respectively. These allow to quantify the scaling behavior of the
average path length with system size as $\langle \ell \rangle\propto L_I^{d_{f}}$, defining the fractal
dimension $d_f$ of the paths, the average path energy $\langle E \rangle \propto L_I^{d_E}$, 
and the fluctuation of the path energies as 
${\rm var}(E)=\langle E^2 \rangle - \langle E \rangle^2 \sim L_I^{2 \Omega}$. 
Note that these energy fluctuations are measured with respect to the 
linear extension of the 
considered lattice graphs. This can be compared
the ground-state energy fluctuations of the DPRM, which
 are commonly measured for polymers
of a fixed length $L$, giving rise to the fluctuation 
exponent $\omega$ defined as
${\rm var}(E_{GS})\propto L^{2 \omega}$, see Ref.\ \cite{derrida1990}. 
Hence, to compare our results with the DPRM case, one should
rewrite the energy fluctuations as a function of the average path length
$\langle \ell \rangle$. This leads to
 a corresponding estimate of $\omega$ 
via using the relation $\omega=\Omega / d_{f}$.

In the following section, we will use the algorithmic procedure outlined above in 
order to study the minimum-energy path problem for an increasing number of iteration
steps $I$ and for a large range of values of the disorder parameter $\rho$.

\section{Results \label{sect:results}}

In the presented study, the disorder distribution is a Gaussian with
mean $\mu$ and  width $\sigma^2=1$. A tunable disorder parameter is
defined as $\rho=1/\mu$, so that  the standard normal distribution is
recovered in the limit $\rho\to \infty$.  Typical values for the
pool-size and iteration steps are $N=10^6$ and $I_{\rm max}=30$,
respectively.

To facilitate intuition, note that as $\mu\to \infty$,
typical minimum-energy paths will exhibit positive
energies. Thus, an increasing path length  will lead to an increasing
path energy. Hence, a minimum-energy path will tend  to be short in
length. Considering hierarchical lattice graphs $G_{I}$, one might
consequently expect a scaling behavior $\langle \ell \rangle\propto
2^I$, implying a scaling exponent $d_f=1$.  On the other hand, as
$\mu\to -\infty$, typical minimum-energy paths will exhibit 
negative energies. Therefore, an increasing path length results in a 
decreasing path energy, leading to expect $\langle \ell \rangle\propto 3^I$,  
and therefore $d_f={\rm log}_2(3)\approx 1.585$.  In between these two
extremal ``trivial'' cases, there exists a particular value
$\rho_c=1/\mu_c$ of the disorder parameter that signifies the onset of
``proliferation'', where $\langle E \rangle=0$ as $I\to
\infty$. Here
 the average path length exhibits a
non-trivial  scaling behavior displayed by a scaling exponent
$1<d_f<1.585$.

At first we attempt to estimate the value of $\rho_c$ by means of the
secant method \cite{num_rec}, considering different initial pools at a given 
value of $I$. Next, we consider one particular initial pool to quantify
the scaling behavior of the average minimum-energy path length and energy. 
Finally, we put under scrutiny the probability density of minimum-energy path energies.

\begin{figure}[t!]
\centerline{\includegraphics[width=1.\linewidth]{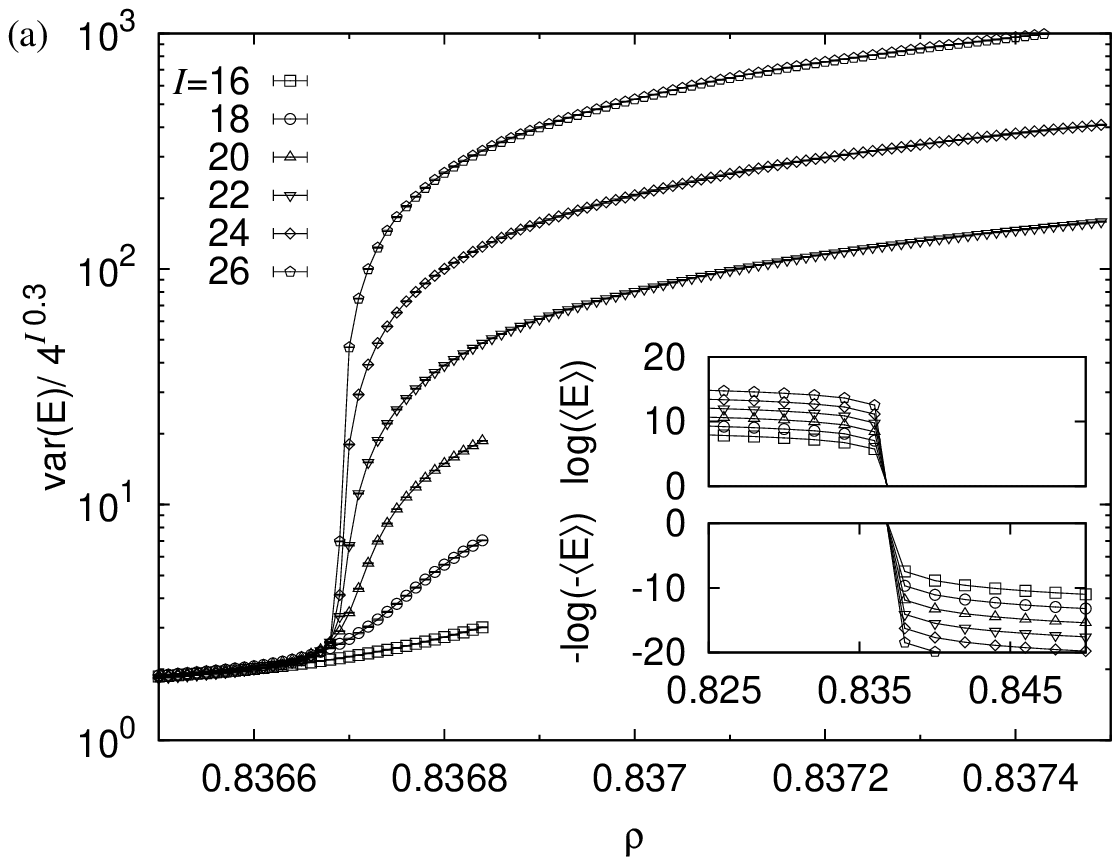}}
\centerline{\includegraphics[width=1.\linewidth]{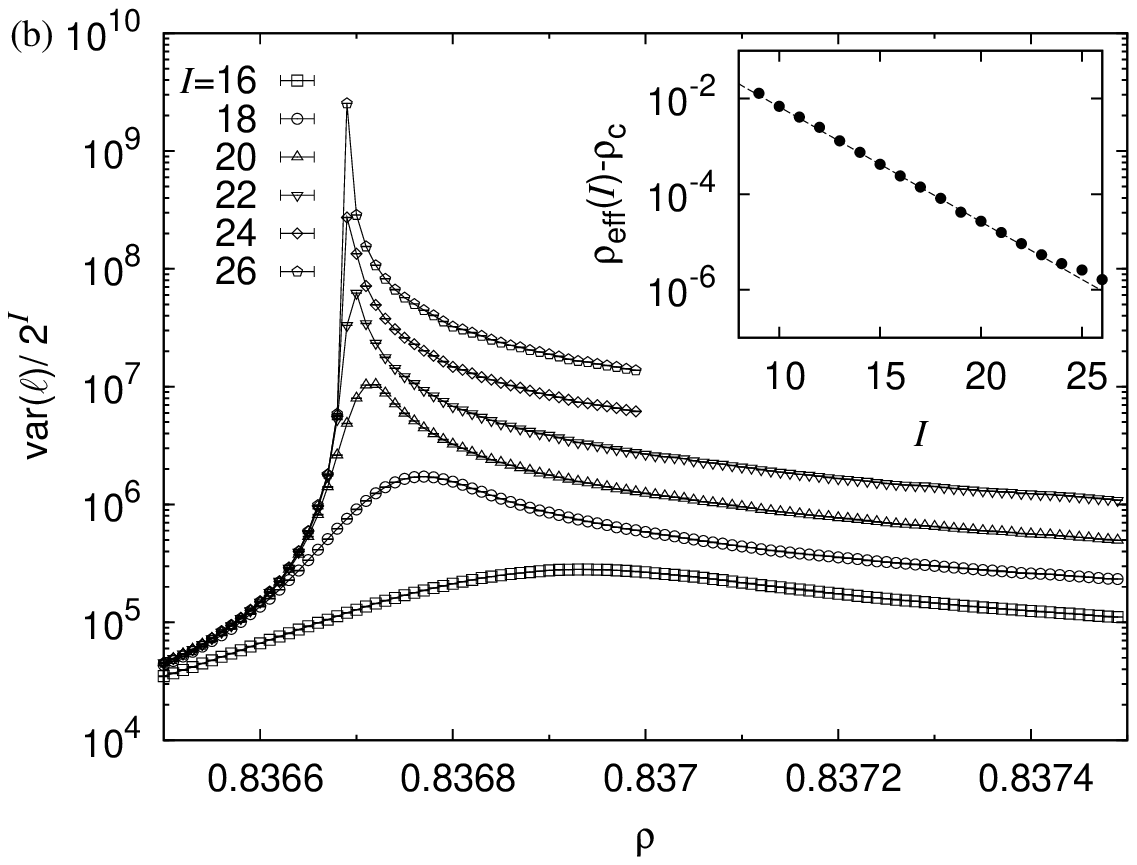}}
\caption{
Variability of the energy and length of the minimum-energy 
paths contained in pools corresponding to different iteration
steps $I$. The data is normalized in a way that curves for 
different values of $I$ overlap as $\rho \to 0$. 
(a) The main plot shows the variance of the path energies,
where the curves indicate a change in the scaling behavior 
at $\rho\approx 0.8367$. This is supported by the scaling
of the average path energy, shown in the inset.
(b) The main plot shows the variability of the path length, 
and the inset illustrates the scaling behavior of the
effective critical points $\rho_{\rm eff}(I)$ that indicate
the associated peak position. 
\label{fig:lenFluct}}
\end{figure}  

\subsection{Location of the critical point where the minimum-energy path energy vanishes}

In order to approximate the critical point $\rho_c$ we considered pools of size $N=10^5$ and $I\leq 22$.
So as to arrive at an estimate of $\rho_c$ at iteration step $I$ we proceeded 
as follows:
Using the secant method we prepared a number of $M=32$ independent estimates of effective critical values
$\rho_{\rm eff}^{(i)}$, where $\langle E \rangle\approx 0$
at the considered value of $I$. The distribution of these
effective critical values (see inset of Fig.\ \ref{fig:secantMethod}) 
is characterized by the average $\rho_c(I)=(1/M)\sum_{i=1}^M \rho_{\rm eff}^{(i)}$.
E.g., at $I=22$ we yield $\rho_{c}(I)=0.8435(7)$, wherein the standard 
deviation among the $M$ independent estimates reads $\sigma_{\rho_c}=0.004$.
The averages exhibit the scaling behavior $\rho_{c}(I) = \rho_c + a 2^{- I b}$, 
where a fit \cite{practicalguide2009}
to the interval $I\in[5,22]$ yields $\rho_c=0.8436(2)$, $a=O(1)$, and $b=0.866(4)$, 
see Fig.\ \ref{fig:secantMethod}. 
The results did not depend much on the pool size, e.g.\ considering $N=10^4$ and 
proceeding as above we find $\rho_c=0.8435(1)$ and $b=0.868(5)$.

\subsection{Trivial to non-trivial transition of the average
minimum-energy path length}
\begin{figure}[t!]
\centerline{\includegraphics[width=1.\linewidth]{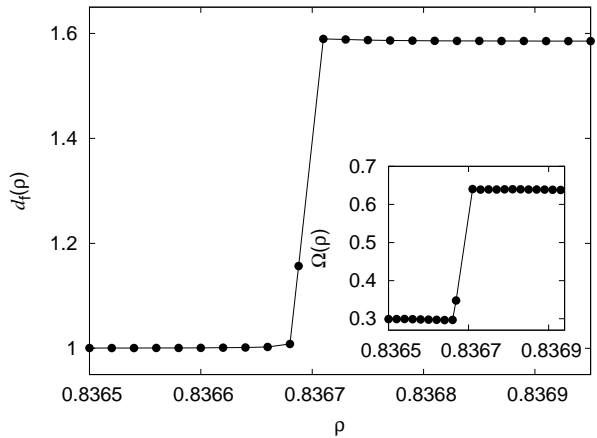}}
\caption{
Critical exponents that characterize the
minimum-energy path length and energy on hierarchical lattice 
graphs for different values of $\rho$ 
below, right at, and above the effective critical point 
$\rho_{\rm eff}=0.836688(1)$ (computed for one exemplary pool of 
size $N=10^6$ considering $I\leq 30$).
The main plot shows the fractal dimension $d_f$ obtained 
from a power law fit to the scaling form $\langle \ell \rangle\propto 2^{I d_f}$
for $I\in [20,30]$ (only exception: the fit at the critical point 
$\rho_c$ was restricted to the interval $I\in [5:10]$).
The inset shows the energy fluctuation exponent $\Omega$ obtained
from a fit to ${\rm var}(E)\propto 2^{2 I \Omega}$ in a similar manner. 
\label{fig:scalingExponents}}
\end{figure}  

As it appears, for large values of $I$ and $\rho_c\approx 0.84$ one should observe a 
vanishing average path energy. In the presented subsection we consider a single pool of size $N=10^6$ and
$I\leq 30$ in order to assess the scaling behavior of the minimum-energy path length and energy.
For that pool we find that the average path energy changes its sign at $\rho\approx0.8367$
(see inset of Fig.\ \ref{fig:lenFluct}(a)). Further, at that approximate value the 
scaling behavior of the fluctuations related to the path energy and length change
significantly (see Figs.\ \ref{fig:lenFluct}(a),(b)).
In this regard, the precise location of the peak position related to 
${\rm var}(\ell)$ (see Fig.\ \ref{fig:lenFluct}(b)) can be used to define an 
iteration-step dependent effective critical point $\rho_{\rm eff}(I)$. 
Seen as a function of $I$, these effective critical points
can be used to pinpoint the precise location where the proliferation 
transition of the path length occurs in the limit $I\to \infty$.
The effective critical values exhibit a scaling of the form
$\rho_{\rm eff}(I) = \rho_c + a 2^{- I b}$, 
where a fit to the interval $I\in[10,20]$ yields the estimates 
$\rho_c=0.836688(1)$, $a=O(1)$, and $b=0.76(2)$, see inset of Fig.\ \ref{fig:lenFluct}(b).
As pointed out above, for $\rho<\rho_c$ an increase in path length most likely results 
in an increasing path energy. Hence, one can expect that for $\rho<\rho_c$ the minimum-energy path
problem investigated here effectively corresponds to the optimal path problem 
studied in Ref.\ \cite{cieplak1995}, wherein a nonegative uniform disorder distribution
was considered. From this it is immediate to expect $d_f=1$, $d_E=1$ and $\Omega=0.3$ for $\rho<\rho_c$.
From a direct fit to the scaling forms $\langle \ell \rangle\propto 2^{I d_f}$, 
$\langle E \rangle \propto 2^{I d_E}$ and 
${\rm var}(E)\propto 2^{2 I \Omega}$ we obtained the numerical values for the scaling exponents 
$d_f$, $d_E$ and $\Omega$ (and consequently the ``corrected'' exponent
$\omega$) as shown in Fig.\ \ref{fig:scalingExponents} and listed in Tab.\ \ref{tab:tab1}.
For values of $\rho$ below and above the critical point, the fits were restricted to 
large iteration steps, i.e.\ $I\in[20,30]$. However, note that it is difficult
to prepare a system right at $\rho_c$: as $I$ increases, fluctuations will eventually cause the 
system to assume the asymptotic scaling behavior characteristic for $\rho<\rho_c$ or $\rho>\rho_c$.
Hence, in order to obtain the scaling exponents for $\rho\approx \rho_c$ the fitting procedure
was restricted to intermediate iteration steps $I\in[5,10]$ ($I\in [5,15]$ in case of $\langle E \rangle$), only. Anyway, for the case of the energy,
we actually require $\langle E \rangle \approx 0$ at $\rho=\rho_c$, 
hence the value
of $d_E$ right at the critical point is of limited relevance and a pure
numerical artifact.

%
\begin{table}[b!]
\caption{\label{tab:tab1}
Critical exponents that characterize the self-affine to self-similar
transition of the minimum-energy path length on hierarchical lattice 
graphs. 
The table lists the 
numerical values of the critical exponents 
below, right at, and above the effective critical point 
$\rho_{\rm eff}=0.836688(1)$, computed for one examplary pool of 
size $N=10^6$ considering $I\leq 30$.
From left to right: 
numerical values of the scaling exponents $d_f$ and $d_E$ for the 
path length and energy, respectively,
the energy fluctuation exponent $\Omega$ (where the energy fluctuations
are considered as a function of the linear extension of the lattice graph),
and the corrected energy fluctuation exponent $\omega$ (where the 
energy fluctuations are considered relative to the actual path length).
} 
\begin{ruledtabular}
\begin{tabular}[c]{l@{\quad}llll}
 & $d_f$ & $d_E$ & $\Omega$ & $\omega$ \\
\hline
$\rho<\rho_c$ 		& 1 & 1 & 0.300(4) & 0.300(4) \\
$\rho\approx\rho_c$ 	& 1.158(1) & 0.3496(2) & 0.347(5) & 0.300(5) \\
$\rho>\rho_c$ 		& 1.5849631(1) & 1.58497(1) & 0.635(5) & 0.401(5)\\
\end{tabular}
\end{ruledtabular}
\end{table}

\begin{figure}[t!]
\centerline{
\includegraphics[width=1.\linewidth]{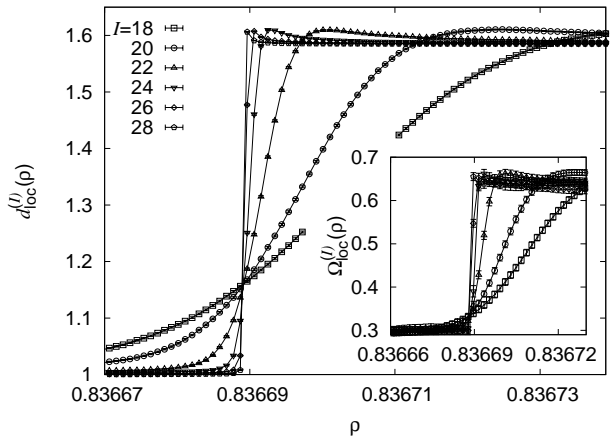}}
\caption{
Local scaling exponents obtained for the length and the energy
fluctuations of the minimum-energy paths, close to the 
critical point $\rho_c=0.836688(1)$. 
The main plot shows the behavior of the local exponents $d_{\rm loc}(\rho)$
associated to the path length
for an increasing number of decimation steps $I$ as a function 
of the disorder parameter $\rho$, and 
the inset illustrates the scaling exponents $\Omega_{\rm loc}(\rho)$ related to the 
fluctuation of the path energies.
\label{fig:locExp}}
\end{figure}

Another means to quantify the scaling behavior of the above observables is given 
by the local scaling exponents. E.g., denoting the average minimum-energy path length 
at a given value of $\rho$ and iteration step $I$ as
$\langle \ell(\rho) \rangle_I$, one can obtain the local analog to the fractal dimension as 
$d_{\rm loc}^{(I)} (\rho) = \log_2(\langle \ell(\rho) \rangle_{I+1}/\langle \ell(\rho) \rangle_{I})$.
The respective error can be obtained via error propagation as 
$\delta d_{\rm loc}^{(I)}= \delta \langle \ell(\rho) \rangle_I / \langle \ell(\rho) \rangle_I + \delta \langle \ell(\rho) \rangle_{I+1} / \langle \ell(\rho) \rangle_{I+1}$. 
The resulting local scaling exponents are shown in Fig.\ \ref{fig:locExp}. 
The local equivalent $\Omega_{\rm loc}^{(I)}$ of the energy fluctuation exponent can 
be computed in similar manner, see inset of Fig.\ \ref{fig:locExp}.
For a given value of $\rho$ and for increasing $I$, the local exponents become independent of $I$
and tend to a limiting value that is within error bars in agreement with the numerical values
of $d_f$ and $\Omega$ listed in Tab.\ \ref{tab:tab1}.

\subsection{Importance sampling results for the ground-state
energy distribution}

\begin{figure}[t!]
\centerline{
\includegraphics[width=1.\linewidth]{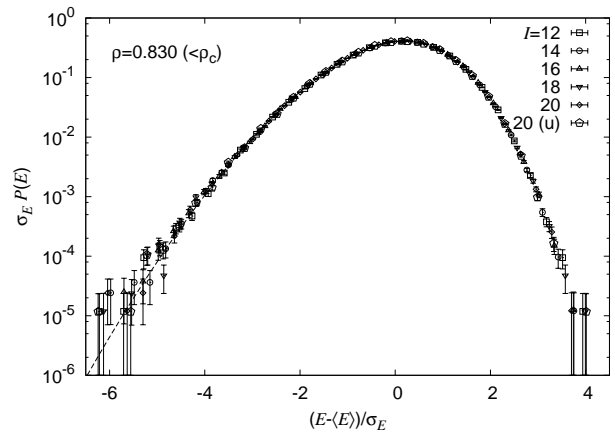}}
\caption{
Simple-sampling estimate of the ground-state energy 
distribution for the minimum-energy path at $\rho=0.830$.
The distribution of the optimal path energy for the case where
edge-energies are drawn uniformly from the interval $[0,1]$
(i.e.\ the case considered in Ref.\ \cite{cieplak1995}) is 
also shown (in the key, the respective symbol is marked as 
(u)).
The dashed line indicates the guiding function fitted to 
the negative tail of the ground-state energy distribution.
\label{fig:ergPdf}}
\end{figure}

As stressed above, for a disorder parameter $\rho<\rho_c$ the
minimum energy path problem considered here effectively corresponds
to the optimal path problem studied in Ref.\ \cite{cieplak1995}.
This is further highlighted by the probability distribution function
(pdf) of minimum-energy
path energies. In this regard, Fig.\ \ref{fig:ergPdf} shows the
simple-sampling estimate
of the ground-state energy distribution for the minimum-energy path 
at $\rho=0.830$ for different iteration steps $I$, obtained from pools of size $N=10^6$.
As evident from Fig.\ \ref{fig:ergPdf}, data 
curves corresponding to different iteration steps $I$ scale according to 
\begin{eqnarray}
P_I(E)=\sigma_E^{-1} {\sf P}( (E-\langle E \rangle)/\sigma_E ),
\end{eqnarray}
where $\epsilon\equiv (E-\langle E \rangle)/\sigma_E $ defines a reduced energy
with $\langle E \rangle$ and $\sigma_E$ describing the average and standard deviation
of the distribution $P_I(E)$, respectively. This means
that the scaling function 
${\sf P}(\epsilon)$ does not depend on the value of $I$.
Note that the rescaled distribution of the optimal path energy for the case were
edge-energies are drawn uniformly from the interval $[0,1]$
(i.e.\ the case considered in Ref.\ \cite{cieplak1995})
assumes the same scaling form and falls onto the same master curve.


Due to the close correspondence between the minimum-energy
path problem studied here and DPRM problem one might expect that the
scaling function ${\sf P}(\epsilon)$ has the shape of a Tracy-Widom 
distribution (see Refs.\ \cite{halpinHealy1995,monthus2006}).
The negative tail of the Tracy-Widom distribution exhibits the
asymptotic scaling $P_{\rm TW}(x) \propto \exp(-c |x|^\eta )$,
where in case of the $d=1$ DPRM (having one space and one time direction) 
one has $\eta_{d=1}^{{\rm DPRM}}=3/2$ exactly \cite{monthus2006}.
The exponent $\eta$, describing the negative tail of the
pdf ${\sf P}(\epsilon)$, is thereby related to the energy fluctuation exponent
$\omega$ (listed in Tab.\ \ref{tab:tab1}) by means of the expression 
$\eta = 1/(1-\omega)$.
Furthermore, SAWs in 
quenched random hierarchical environments at the critical point were 
studied \cite{Doussal1991} using real-space renormalization techniques 
(quite similar to the approach presented here). Also in that case
both tails of the scaling function ${\sf P}(\epsilon)$
are consistent with an exponential decay as given above. 
Under the assumption that 
the tails of the distribution reproduce under rescaling, they arrive 
at the estimates $\eta_-=1/(1-\Omega/d_f)$ and $\eta_+ = 1/(1-\Omega)$
for the negative and positive tail, respectively.
However, their numerical results then did not allow to conclude with 
precise numerical estimates for the exponents $\eta_{\pm}$. 

Subsequently we address the question whether the 
distribution of ground-state energies in the minimum-energy
path problem for $\rho<\rho_c$, $\rho\approx \rho_c$, and $\rho>\rho_c$
is consistent with a Tracy-Widom scaling form and we attempt
to obtain a numerically precise estimate of the negative tail exponent $\eta$
for the above three cases.
To this end, we consider an
importance sampling procedure in the disorder 
\cite{align2002,rare-graphs2004}, where the 
sampling process is controlled by a guiding function \cite{koerner2006}.
This allows to compute the negative-energy tails of the ground-state energy 
distribution up to $12$ standard deviations away from its mean.
Similar to Ref.\ \cite{monthus2006} we consider the negative tails of the 
distribution for the reduced energy $\epsilon \leq 1$ only.
We further use a guiding function 
\begin{equation}
G(\epsilon)=\exp(a-b|\epsilon+c|^\eta)
\end{equation}
 in order to estimate the parameters
that characterize best the simple sampling distributions at the three 
values
$\rho=0.83,\, \rho=\rho_c,\, \rho=0.86$. 
The respective estimates are listed in Tab.\ \ref{tab:tab2} and the 
guiding function for $\rho=0.83$ is indicated as dashed line in Fig.\ \ref{fig:ergPdf}.

%
\begin{table}[b!]
\caption{\label{tab:tab2}
Parameters for the guiding function used during the 
importance sampling procedure for the listed values $\rho$
of the disorder parameter. The values are obtained 
by fitting the function $G(\epsilon)$ (see text) to the negative
tail of the distribution of the path-energies at
iteration level $I$. 
} 
\begin{ruledtabular}
\begin{tabular}[c]{l@{\quad}llllll}
$\rho$ & $I$ & $a$ & $b$ & $c$ & $\eta$ & $\chi^2/{\rm dof}$ \\
\hline
$0.83$ 	 & $20$ & $-0.98(4)$ & $0.72(5)$ & $0.14(8)$ & $1.55(3)$ & $1.01$ \\ 
$\rho_c$ & $15$ & $-1.00(3)$ & $0.81(6)$ & $0.24(7)$ & $1.49(4)$ & $0.81$ \\
$0.86$ 	 & $20$ & $-0.94(2)$ & $0.56(4)$ & $0.03(7)$ & $1.84(4)$ & $0.78$
\end{tabular}
\end{ruledtabular}
\end{table}

Let ${\sf P}(\epsilon)$ describe the true probability density function
of observing a minimum-energy path with reduced energy $\epsilon$ for
the model under consideration (bear in mind  that it holds that ${\sf
P}(\epsilon)=\sigma_E P_I(E)$, where $\epsilon\equiv (E - \langle E
\rangle)/\sigma_E$).  In order to arrive at an ImSa
estimate ${\sf P^{\rm IS}}(\epsilon)$ that approximates ${\sf P}(\epsilon)$,
we divide the generation of the $I$ iterations into
two parts, consisting of $I-\Delta I$ and $\Delta I$ iterations, 
where $\Delta I$ is small, we consider
$\Delta I=2, 3$ or $4$. 
The generation of the first $I-\Delta I$ iterations,
is performed in the usual simple-sampling way,
leading to a large ($N=10^6$) pool $\mathcal{P}_{I-\Delta I}$.
The final $\Delta I$ iterations should be done in a way that also
the interesting tails of the distribution are sampled. For this
purpose, we generate a \emph{Markov chain} 
 $\mathcal{P}_{0}^{\rm IS}$ $\to$  $\mathcal{P}_{1}^{\rm IS}$
$\to$ $\mathcal{P}_{2}^{\rm IS}$ $\to$ $\ldots$ of pools, which 
all are subsets of $\mathcal{P}_{I-\Delta I}$, but the sampling
is done in a way that also the tails of the distribution
of reduced energy values are sampled.

The initial pool $\mathcal{P}_0^{\rm IS}$ is created by picking 
a (uniformly sampled) random subset of $5^{\Delta I}$ edges from    
$\mathcal{P}_{I-\Delta I}$. For this pool now the final $\Delta I$
levels of the hierarchy are performed within one graph:
The $5^{\Delta I}$ edges comprising
the sampling pool can be arranged  into one particular realization of
a hierarchical graph $G_{\Delta I}$.  
Upon stepwise decimation $G_{\Delta I} \to
G_0$ this  yields one particular edge with 
an edge energy $E_0$ and a corresponding
 reduced energy
$\epsilon_0\equiv (E_0^{\rm IS} - \langle E
\rangle)/\sigma_E$, wherein $\langle E \rangle$ and $\sigma_E$
describe the simple sampling estimate of $P_I(E)$. Note that 
the probability in the tails is very small, hence ImSa will
not change  $\langle E \rangle$ and $\sigma_E$ considerably.

The Markov chain Monte Carlo step reads as follows:
From a given
sampling pool $\mathcal{P}_{i}^{\rm IS}$ we construct the next
sampling pool $\mathcal{P}_{i+1}^{\rm IS}$ using  the following 2-step
procedure:

\begin{enumerate}
\item[(1)] randomly choose a fraction $p$ of edges contained in the sampling pool 
$\mathcal{P}_{i}^{\rm IS}$
and replace those edges by new edges chosen from the large pool $\mathcal{P}_{I-\Delta I}$. This
then specifies a candidate $\mathcal{P}^{\prime}$ for the next sampling pool, characterized
by the reduced energy $\epsilon^\prime$.
\item[(2)] set $\mathcal{P}_{i+1}^{\rm IS} = \mathcal{P}^{\prime}$ with 
probability
\begin{eqnarray}
P_{\rm accept} = {\rm min}\Big[ \frac{G(\epsilon_i)}{G( \epsilon^\prime )},1 \Big].
\end{eqnarray}
Set $\mathcal{P}_{i+1}^{\rm IS} = \mathcal{P}_{i}^{\rm IS}$ otherwise.
\end{enumerate}
To complete the importance sampling simulation, the evolution of the initial 
sampling pool $\mathcal{P}_0^{\rm IS}$ is followed a number of $M$ steps.
The resulting $M+1$ reduced energy values $\epsilon_0 \ldots \epsilon_M$ comprise an auxiliary 
distribution ${\sf P}^{\rm IS}(\epsilon)$ that describes the probability by means of which 
a reduced path energy $\epsilon$ is visited within the ImSa procedure.
Since the importance sampling is designed such that a sampling pool having reduced energy $\epsilon$ 
is encountered with probability $\propto 1/G(\epsilon)$, the distribution ${\sf P}^{\rm IS}(\epsilon)$
 is further given by the ratio
${\sf P}^{\rm IS}(\epsilon)={\sf P}(\epsilon)/G(\epsilon)$ (where ${\sf P}(\epsilon)$ 
describes the true pdf of observing a minimum-energy path with reduced energy $\epsilon$
for the considered model system).
As long as the guiding function $G(\epsilon)$ provides a reasonable approximation to the
true distribution of path energies, the auxiliary distribution ${\sf P}^{\rm IS}(\epsilon)$ obtained 
using the IS procedure is rather ``flat''. Thus, regarding the target distribution ${\sf P}(\epsilon)$ one might hope to improve on the negative-tail
statistics provided by a simple sampling approach.
As discussed in Ref.\ \cite{koerner2006}, successive configurations (i.e.\ sampling pools)
encountered during an ImSa simulation are not independent. 
As a remedy one might consider the autocorrelation function 
\begin{eqnarray}
\chi(\Delta i)=\frac{\langle E_i E_{i+\Delta i} \rangle - \langle E_i \rangle \langle E_{i+\Delta i}\rangle }{\langle E^2_i\rangle - \langle E_i\rangle^2}
\end{eqnarray}
 associated to the sequence of energy values $E_i$ obtained from the 
importance sampling procedure. The number of Monte Carlo steps that have to elapse
until the autocorrelation function decays to $1/e$ gives the respective autocorrelation time $\tau_E$.
Sampling pools that are separated by $\approx \tau_E$ Monte Carlo steps can 
be considered effectively uncorrelated. Finally, the truncated sequence of effectively uncorrelated 
energy values can be analyzed similar to the simple-sampling data.
\begin{figure}[t!]
\centerline{
\includegraphics[width=1.\linewidth]{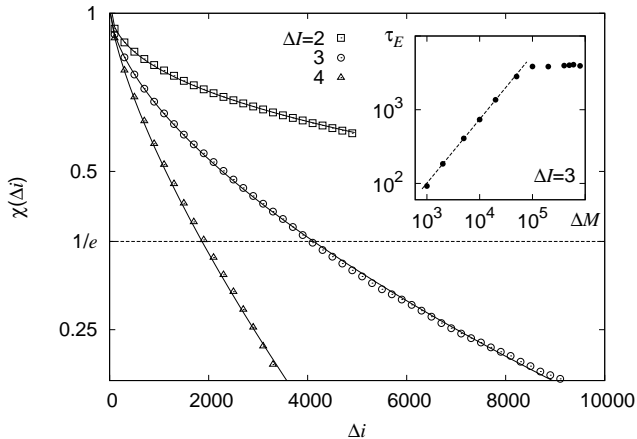}}
\caption{
The main plot shows the autocorrelation functions $\chi(\Delta i)$ for 
importance sampling simulations at $\rho=0.83$ and $\Delta I=2,3,4$.
The solid lines indicate a best fit of the respective data to 
a stretched exponential function $\chi(\Delta i)\propto \exp\{-(\Delta i/\tau_E)^\beta\}$.
The resulting fit-parameters are listed in the text.
The inset indicates the effective autocorrelation times that 
result from an analysis of slices of $\Delta M$ successive energy values
obtained during the ImSa simulation at $\Delta I=3$. For $\Delta M<10^5$ 
the autocorrelation time appears to increase $\propto \Delta M^{0.87(3)}$ 
(the corresponding fit is indicated as a dashed line).
\label{fig:acTime}}
\end{figure}  

The results discussed below were obtained for the choice $p=0.2$, where we restricted the IS 
procedure to $-12\leq \epsilon \leq -1$. We further performed a number of $M=10^7$ Monte Carlo 
steps to estimate the distribution of path energies in a target pool corresponding to $I=20$.
In order to assess the autocorrelation time we considered IS simulations at a disorder parameter
$\rho=0.83$ and for $\Delta I=2,3$ and $4$.
As it appears, the autocorrelation function can be well described by a stretched exponential,
i.e.\ $\chi(\Delta i)\propto \exp\{-(\Delta i/\tau_E)^\beta\}$, where $\beta<1$. 
A best fit of that relation to the data yields the parameters
$\tau_E\approx 18677, 4023, 1817$ and $\beta\approx 0.41,0.61,0.72$ for $\Delta I=2,3,4$, respectively.
Such a stretched exponential might result from a continuous sum of pure exponential decays \cite{johnston2006}.
Further, a stretched exponential decay of the energy autocorrelation function 
was, above the respective critical temperature, also observed for the
$2D$ and $3D$ fully frustrated Ising model \cite{Franzese1998}.
To check whether there are different autocorrelation times relevant on different 
timescales of the ImSa simulation we performed the following analysis:
we subdivided the ``long'' simulation run at $\Delta I=3$ into $m=M/\Delta M$ ``shorter'' runs, each of length $\Delta M$.
The $m$ runs of length $\Delta M$ are then considered as being independent and the characteristic 
autocorrelation time $\tau_E(\Delta M)$ for the shorter sequences is estimated. 
In this regard, for $\Delta M \leq 5 \times 10^3$ it was sufficient to consider a pure power 
law fit-function. For values of $\Delta M$ larger than that, a stretched exponential turned out to be more adequate.
As shown in the inset of Fig.\ \ref{fig:acTime}, we found that for $\Delta M < 10^5$ the effective autocorrelation 
time is well described by an algebraic dependence $\tau_E(\Delta M)= 0.24(8)\times \Delta M^{0.87(3)}$, 
whereas for $\Delta M>10^5$ the value of $\tau_E$ did not increase further, i.e.\
we observed $\tau_E(\Delta M>10^5)\approx 4000$.
As pointed out above, sampling pools that are separated by more than $\tau_E \approx 4000$ Monte Carlo steps
are effectively uncorrelated. One might now argue that, in order to use only 
uncorrelated values of $E$ to construct the distribution ${\sf P}^{\rm IS}(\epsilon)$ (and hence ${\sf} P(\epsilon)$), 
one should keep only every $\tau_E$-th energy value.
However, in Ref.\ \cite{monthus2006} the authors concluded that if during the $M$ Monte Carlo steps the 
interval $[\epsilon_{\rm min},\epsilon_{\rm max}]$ is crossed sufficiently often,
it is not necessary to discard any energy values obtained during the IS simulation.
Here, for $\Delta I=3$, considering the interval $[-12,-1]$ and performing $M=10^7$ Monte 
Carlo steps at $\rho=0.83$ we found a number of $n_{\rm cross}=366$ interval crossings.
In agreement with Ref.\ \cite{monthus2006} we observed that it makes no difference whether 
the sequence of energy values collected during the IS procedure was truncated or not, 
the resulting pdf ${\sf P}(\epsilon)$ remained almost unchanged (apart from effects due to 
the different sample sizes used to construct the pdfs).

\begin{figure}[t!]
\begin{center}
\includegraphics[width=1.\linewidth]{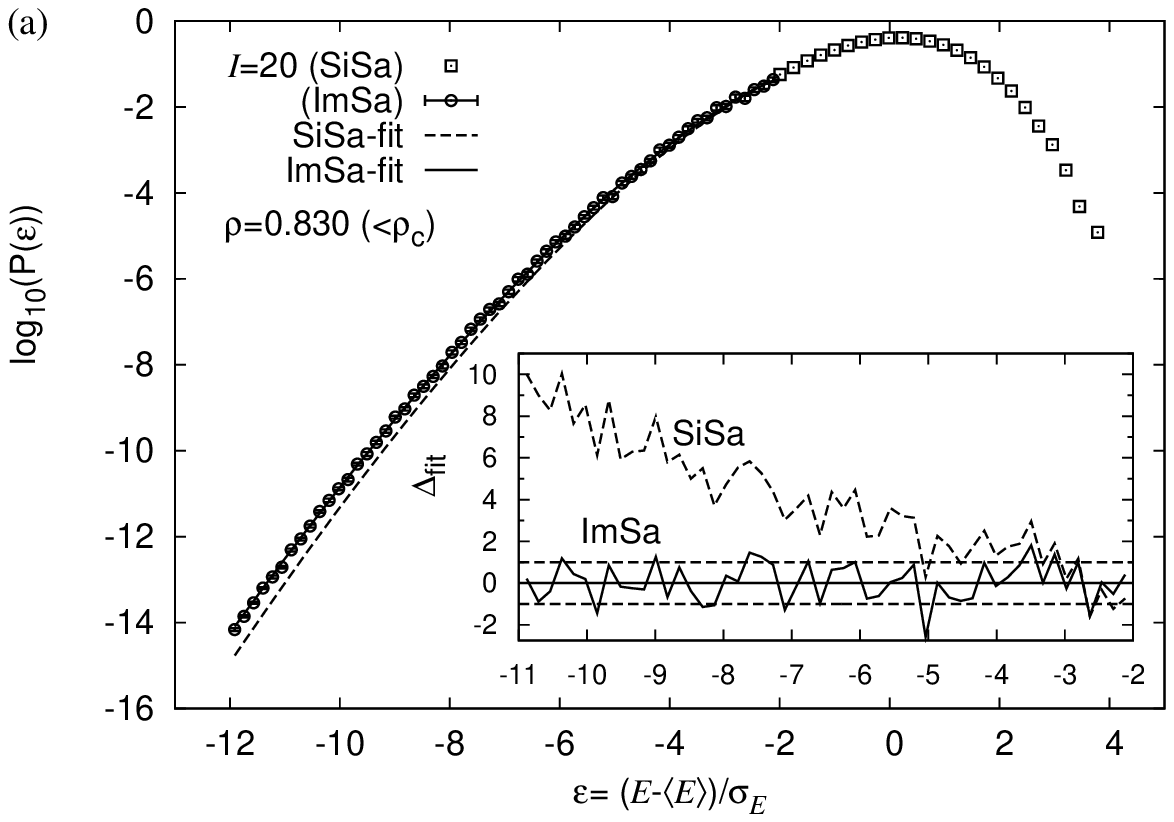}
\includegraphics[width=1.\linewidth]{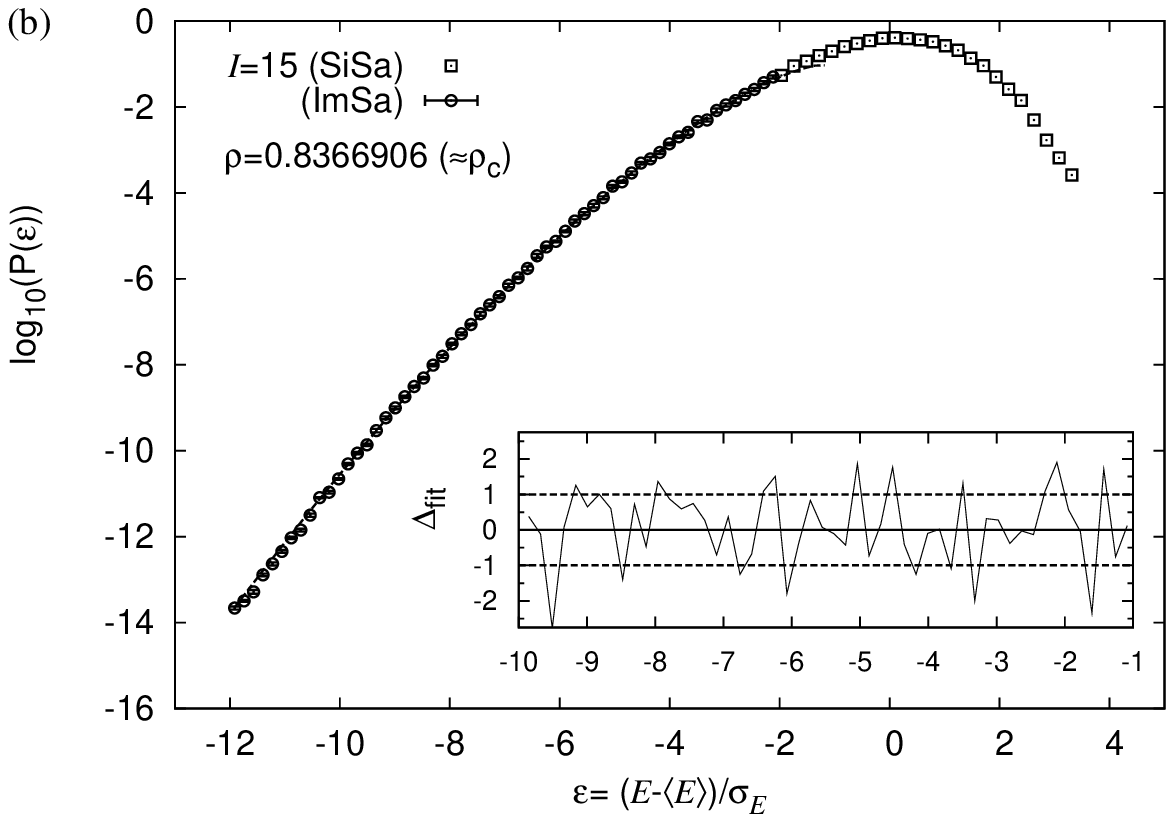}
\includegraphics[width=1.\linewidth]{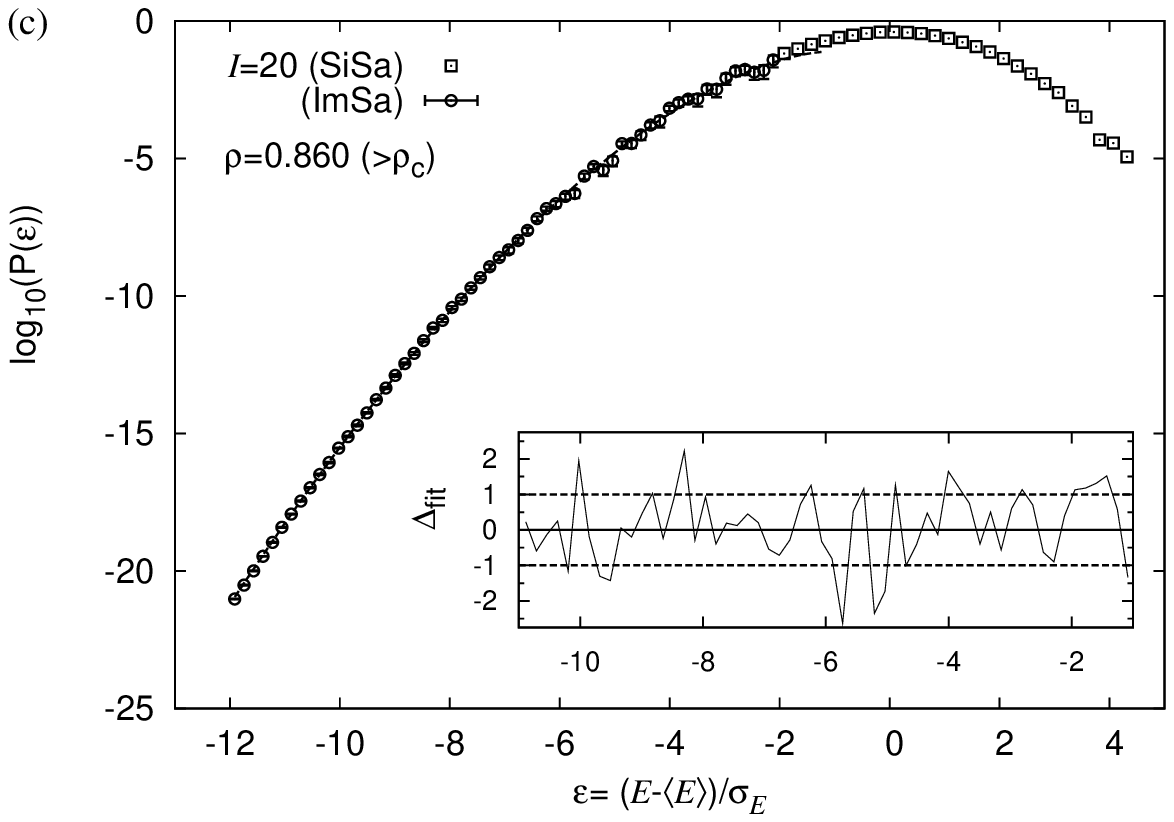}
\end{center}
\caption{
Pdf of observing a minimum-energy path with reduced energy 
$\epsilon$, obtained by an importance sampling Monte Carlo
simulation in the disorder.
The main plots show a semi-logarithmic plot of the distribution
${\sf P}(\epsilon)$ at (a) $\rho=0.83$, (b) $\rho\approx \rho_c$,
and (c) $\rho=0.86$. Data points at $\epsilon < -2$ ($\geq-2$) refer
to the ImSa (SiSa) estimate of the pdf. In either case the error bars
are of the order of the symbol size.   
The insets show the normalized deviation $\Delta_{\rm fit}$ of a best 
fit for $G(\epsilon)=\exp\{a-b |x+c|^\eta\}$ to the negative tail.
The respective parameters $\eta$ are listed in Tab.\ \ref{tab:tab3}. 
\label{fig:impSamp}}
\end{figure}  

%
\begin{table}[b!]
\caption{\label{tab:tab3}
Exponent $\eta$, describing the scaling of the negative
tail of the pdf ${\sf P}(\epsilon)$, obtained by fitting
the function $G(\epsilon)$ (see text) to the data resulting
from the ImSa procedure. From left to right: value $\rho$ of
the disorder parameter, interval over which the fit was 
performed, exponent $\eta$ (as well
as $1-1/\eta$) and
reduced chi-square $\chi^2/{\rm dof}$.
} 
\begin{ruledtabular}
\begin{tabular}[c]{l@{\quad}llll}
$\rho$ & $[\epsilon_{-},\epsilon_{+}]$& $\eta$  & $1-1/\eta$ & $\chi^2/{\rm dof}$ \\
\hline
$0.83$	 & $[-11,-1.5]$ & $1.42(3)$& $0.30(1)$ & $0.88$\\
$\rho_c$ & $[-10,-1]$   & $1.42(2)$& $0.30(1)$ & $1.21$\\
$0.86$ 	 & $[-12,-1]$   & $1.65(6)$& $0.39(2)$ & $1.10$
\end{tabular}
\end{ruledtabular}
\end{table}

Below we present the results obtained for ImSa simulations at $\rho=0.83, \rho_c, 0.86$
considering $\Delta I=3$ and a target distribution at $I=20$ (only the simulation at $\rho_c$ 
was carried out for a target pool at $I=15$).
Once the distribution ${\sf P}^{\rm IS}(\epsilon)$ is obtained from the 
ImSa procedure, it can immediately be transformed to the desired pdf ${\sf P}(\epsilon)$.
A comparison of the SiSa and ImSa pdf shows 
that the absolute probabilities in the overlapping region $\epsilon \in [-6,-1]$ do not 
coincide. This is due to the restriction of the relative energies to 
the interval $\epsilon\in [-12,-1]$ during the ImSa simulation. One can 
easily account for this discrepancy by requiring that the SiSa and ImSa
estimates coincide in the overlapping region, and by rescaling the ImSa estimate accordingly.
In Figs.\ \ref{fig:impSamp}(a-c), the resulting 
pdfs of observing a minimum-energy path with reduced energy 
$\epsilon$ for $\rho=0.83,\rho_c,0.86$ are shown, respectively.
Note that the pdfs are represented using histograms that consist of $64$ bins, each.
In either case, the ImSa (SiSa) estimate is depicted for relative energies $\epsilon<-2$ ($\geq -2$).
As evident from the figures, using the ImSa procedure probabilities as small 
as ${\sf P}(\epsilon)\propto 10^{-20}$ can be reached, in contrast to $\propto 10^{-6}$ for 
a SiSa approach, cf.\ Fig.\ \ref{fig:ergPdf}.
The insets to Figs.\ \ref{fig:impSamp}(a-c) indicate the normalized deviation of a best fit of the 
function $G(\epsilon)=\exp\{a-b|\epsilon-c|^\eta\}$ to the negative tail of ${\sf P}(\epsilon)$.
Once the fit is performed the deviation is obtained as 
$\Delta_{\rm fit}(\epsilon)= (P(\epsilon)-G(\epsilon))/\Delta P(\epsilon)$,
where $\Delta P(\epsilon)$ indicates the measurement error on $P(\epsilon)$ as obtained by 
bootstrap resampling \cite{practicalguide2009}. 
The observation that $\Delta_{\rm fit}$ (considering a fit of $G(\epsilon)$ to the ImSa data) 
is of order one and changes
sign in an irregular fashion indicates that there are no systematic deviations and that 
the fit-function $G(\epsilon)$ represents a proper approximation to the negative tail of the 
observed pdf.
In particular, Fig.\ \ref{fig:impSamp}(a) highlights that a fit 
to the SiSa pdf might be misleading if one is interested in the true scaling behavior
of ${\sf P}(\epsilon)$ as $\epsilon\to -\infty$. For that purpose, the above fit-function
with fitting parameters obtained for the SiSa data (dashed curve in the main plot) was 
used to compute the normalized deviation to the ImSa data (dashed line in the inset). 
Referring to this one finds rather strong systematic deviations where, e.g., $\Delta_{\rm fit}(-11)\approx 10$.
The parameters $\eta$ that correspond to a best fit to the ImSa data are listed in Tab.\ \ref{tab:tab3}.

As pointed out above, previous studies suggested that the exponent $\eta$ 
is related to the energy fluctuation exponent
$\omega$ (listed in Tab.\ \ref{tab:tab1}) by means of the expression 
$\omega = 1-1/\eta$.
Here, for the minimum-energy path problem on hierarchical lattice 
graphs we find that this expression holds for all values
of $\rho$ thus considered. To facilitate comparison, the values 
$1-1/\eta$ are listed in Tab.\ \ref{tab:tab3}. 


\section{Conclusions \label{sect:conclusions}}

In the presented article we have investigated a particular MWP problem on 
MK hierarchical graphs. It is quite similar to earlier polymer \cite{kardar1987,Derrida1989,derrida1990,Doussal1991,Devillard1993,Cao1993,monthus2008} and
optimal path problems \cite{cieplak1994,cieplak1995,Hansen2004,buldyrev2006}.
The important difference is that for a considerable fraction
of negative edge energies, the total energy of a path may be reduced
by taking longer paths (which leads to a different universality class).
In the same fashion as the optimal path problem on hierarchical graphs, studied in 
Ref.\ \cite{cieplak1995}, corresponds to the (generic) non-directed optimal path 
problem \cite{schwartz1998}, the minimum-energy path problem studied here corresponds
to the negative-energy percolation problem \cite{melchert2008} in which there is 
a path forced onto the system and where the disorder is ``weak'' enough to render
the appearance of loops irrelevant for the scaling behavior of the path.

Here, the scaling properties of the MWP obtained after the decimation of huge hierarchical graphs 
change with increasing edge-disorder, leading from a phase where the 
path displays a self-affine scaling behavior to a phase where the path displays
a (statistically) self-similar scaling behavior. We characterized the respective phase transition
by monitoring the length and energy of the MWPs as function of the disorder and
quantified the scaling behavior of the observables (and their fluctuations) by
means of proper critical exponents.
While the scaling of the observables off criticality can be explained intuitively, 
the scaling behavior found at the critical point of the model is nontrivial and
compares well to the scaling observed for the optimal path problem on the same hierarchical
lattice in the limit of ``strong'' disorder \cite{cieplak1995}. 
However, note that the precise optimization criteria of both models are slightly different:
while in the optimal path problem \cite{cieplak1995,schwartz1998} one aims to minimize the largest energy 
along a sub-path, here one strives after minimizing the sum of energies along a sub-path.
Note that this was already realized for the respective models on regular lattice
graphs, where quite similar scaling exponents for the average path length were found 
in dimensions $d=2$ through $6$, see Ref.\ \cite{melchert2010a}.

Further, we performed an importance sampling simulation for the ground-state
energy distribution of the paths and confirmed that it is consistent with a 
Tracy-Widom scaling form, similar to the directed polymer in a random medium \cite{monthus2006}).
Using the importance sampling procedure allowed for an analysis of the ground-state
energy distribution down to probabilities as small as $\propto 10^{-20}$ 
(in contrast, a convenient simple sampling approach only allows to 
reach $\propto 10^{-6}$), i.e.\ up to $12$ standard deviations away from 
the mean of the respective distribution.
This leads to a precise estimate of the scaling behavior of the negative
tail of the ground-state energy distribution. For all values of the disorder parameter
considered here, the respective exponents $\eta$ could be related to the energy-fluctuation 
exponents $\omega$ via the relation $\omega=1-1/\eta$.


\begin{acknowledgments}
OM acknowledges financial support from the DFG (\emph{Deutsche Forschungsgemeinschaft})
under grant HA3169/3-1.
The simulations were performed at the HPC Cluster HERO, located at 
the University of Oldenburg (Germany) and funded by the DFG through
its Major Instrumentation Programme (INST 184/108-1 FUGG) and the
Ministry of Science and Culture (MWK) of the Lower Saxony State.
\end{acknowledgments}


\bibliography{lit_nwp_MK_impSamp.bib}

\end{document}